\documentclass[a4paper,10pt,twocolumn]{aastex631}

\hypersetup{linkcolor=red,citecolor=blue,filecolor=cyan,urlcolor=magenta}
\usepackage{natbib}
\usepackage{amsthm}
\usepackage{amssymb}
\usepackage{amsmath}
\usepackage{bm}
\usepackage[utf8]{inputenc}
\usepackage{graphicx}
\usepackage[caption=false]{subfig}
\usepackage{float}
\usepackage{sidecap}

\submitjournal{ApJ}

\shorttitle{2D--PIC simulations of firehose instabilities}
\shortauthors{R.A. L\'opez et al.}

\begin{document}

\title{Mixing the solar wind proton and electron scales. Theory and 2D--PIC simulations of firehose instability}

\correspondingauthor{R.A. L\'opez}
\email{rlopez186@gmail.com}

\author[0000-0003-3223-1498]{R. A. L\'{o}pez}
\affiliation{Departamento de F\'{\i}sica, Facultad de Ciencias, Universidad de Santiago de Chile, Santiago, Chile}

\author[0000-0001-9293-174X]{A. Micera}
\affiliation{Solar-Terrestrial Centre of Excellence - SIDC, Royal Observatory of Belgium, Brussels, Belgium}

\author[0000-0002-8508-5466]{M. Lazar}
\affiliation{Centre for Mathematical Plasma Astrophysics, Department of Mathematics, KU Leuven, Celestijnenlaan 200B, B-3001 Leuven, Belgium}
\affiliation{Institute for Theoretical Physics IV, Faculty for Physics and Astronomy, Ruhr-University Bochum, D-44780 Bochum, Germany}

\author[0000-0002-1743-0651]{S. Poedts}
\affiliation{Centre for Mathematical Plasma Astrophysics, Department of Mathematics, KU Leuven, Celestijnenlaan 200B, B-3001 Leuven, Belgium}
\affiliation{Institute of Physics, University of Maria Curie-Sk{\l}odowska, Pl.\ M.\ Curie-Sk{\l}odowska 5, 20-031 Lublin, Poland}

\author[0000-0002-3123-4024]{G. Lapenta}
\affiliation{Centre for Mathematical Plasma Astrophysics, Department of Mathematics, KU Leuven, Celestijnenlaan 200B, B-3001 Leuven, Belgium}

\author[0000-0002-2542-9810]{A. N. Zhukov}
\affiliation{Solar-Terrestrial Centre of Excellence - SIDC, Royal Observatory of Belgium, Brussels, Belgium}
\affiliation{Skobeltsyn Institute of Nuclear Physics, Moscow State University, Moscow, Russia}

\author[0000-0003-1970-6794]{E. Boella}
\affiliation{Physics Department, Lancaster University, Lancaster, UK}
\affiliation{Cockcroft Institute, Daresbury Laboratory, Warrington, UK}

\author[0000-0003-0465-598X]{S. M. Shaaban}
\affiliation{Theoretical Physics Research Group, Physics Department, Mansoura University, Mansoura, Egypt}

\begin{abstract}
Firehose-like instabilities (FIs) are cited in multiple astrophysical applications. Of particular interest are the kinetic manifestations in weakly-collisional or even collisionless plasmas, where these instabilities are expected to contribute to the evolution of macroscopic parameters.
Relatively recent studies have initiated a realistic description of FIs, as induced by the interplay of both species, electrons and protons, dominant in the solar wind plasma. 
This work complements the current knowledge with new insights from linear theory and the first disclosures from 2D PIC simulations, identifying the fastest growing modes near the instability thresholds and their long-run consequences on the anisotropic distributions. 
Thus, unlike previous setups, these conditions are favorable to those aperiodic branches that propagate obliquely to the uniform magnetic field, with (maximum) growth rates higher than periodic, quasi-parallel modes.
Theoretical predictions are, in general, confirmed by the simulations. The aperiodic electron FI (a-EFI) remains unaffected by the proton anisotropy, and saturates rapidly at low-level fluctuations.
Regarding the firehose instability at proton scales, we see a stronger competition between the periodic and aperiodic branches. For the parameters chosen in our analysis, the a-PFI is excited before than the p-PFI, with the latter reaching a significantly higher fluctuation power. However, both branches are significantly enhanced by the presence of anisotropic electrons. 
The interplay between EFIs and PFIs also produces a more pronounced proton isotropization.

\end{abstract}

%
\section{Introduction}\label{sec:intro}
%
The solar wind has an impressive ability to self-control the physical properties of plasma particle populations by redistributing the energy
from large to small scales, between different species of plasma particles, e.g., protons and electrons \citep{Marsch-2006, Verscharen2019}.
At large enough heliocentric distances (e.g., 1~AU and beyond), the hot and dilute plasmas become collision-poor or even collision-less, and the major role in the energy
exchange between plasma populations is expected to be played by the waves and fluctuations, especially those characteristic to proton and electron scales \citep{Pierrard-etal-2011, MArsch-2012}. These
fluctuations are locally enhanced by the kinetic instabilities, driven by the velocity space anisotropy of plasma particles \citep{Wilson-etal-2013, Gary-etal-2016, Woodham-etal-2019, Hunana2019, Hunana2019b}. In the absence of collisions, these instabilities can self-consistently constrain the temperature anisotropy of plasma particles, as, for instance, in the expanding solar wind, where firehose instabilities are believed to induce particle heating (cooling) in the direction perpendicular (parallel) to the magnetic field ${\bf B}_0$ and restore isotropy~\citep{Stverak-etal-2008, Bale-etal-2009}.  

The primary, basic characterization of an instability triggered by a certain anisotropic plasma population is usually made by minimizing the effects of other populations present, considering them isotropic.
Thus, for isotropic ions/protons, but anisotropic electrons (subscript $e$ in the following) with $A_e = T_{e,\perp}/ T_{e,\parallel} < 1$, where $\parallel, \perp$ signify gyrotropic directions with respect to the local magnetic field, linear kinetic theory predicts two branches of electron firehose instability (EFI). 
While the periodic EFI (p-EFI) mode with finite frequency ($\omega \ne 0$) is obtained for parallel direction and extends to small oblique angles of propagation \citep{Paesold-Benz-1999, Li-Habbal-2000, Camporeale-Burgess-2008}, the aperiodic EFI (a-EFI) branch is purely growing ($\omega = 0$) and should develop only for oblique angles (${\bf k} \times {\bf B}_0\ne 0$, with (maximum) growth rates higher than those of p-EFI \citep{Li-Habbal-2000, Camporeale-Burgess-2008, Shaaban-etal-2019a}.
Numerical simulations confirm these predictions showing highly oblique a-EFI dominating the initial growth, although fluctuations are obtained at all propagation angles \citep{Camporeale-Burgess-2008,Lopez-etal-2019,Innocenti2019,Micera2021}.
On the other hand, if electrons are assumed isotropic and only protons (subscript $p$ in the following) exhibit anisotropic temperature, with $A_p = T_{p,\perp} / T_{p,\parallel} <1$, linear theory predicts a similar dual spectra of proton firehose instabilities (PFIs) \citep{Yoon-etal-1993, Hellinger-Matsumoto-2000, Shaaban-etal-2021}. 
However, in this case, the periodic PFI [p-PFI, also known as whistler FI due to the right-hand polarization~\citep{Hellinger-Matsumoto-2000}] is more competitive. It shows growth rates exceeding those of aperiodic PFI (a-PFI) for an extended range of parameters in a plane of proton parallel beta $\beta_{p,\parallel} = 8 \pi n k_B T_{p,\parallel}/B_0^2$ vs proton anisotropy $A_p$, see plate 3 in \cite{Hellinger-Matsumoto-2000}. The growth rates of a-PFI become superior only for a relatively narrow parametric domain near the instability thresholds, i.e., from large deviations from isotropy ($A_p << 1$) and relatively small $\beta_{p,\parallel} \gtrsim 1$, up to small deviations from isotropy ($A_p \lesssim 1$) and large $\beta_{p,\parallel} >1$  \citep{Hellinger-Matsumoto-2000} [see also some of our comparative graphics in section \ref{theory}]. Numerical simulations confirm an extended dominance of p-PFI, that, however, saturates rapidly and is followed by the growth of a-PFI with a subsequent significant contribution to the relaxation of protons \citep{Gary-etal-1998, Hellinger-Matsumoto-2001}.                      

Electrons and protons are the dominant plasma species in the solar wind. Investigations of their mutual effects on the excitation of firehose instabilities have also been initiated relatively recently, providing the first insights from linear theory \citep{Michno-etal-2014, Maneva-etal-2016,Shaaban-etal-2017}, quasi-linear theory~\citep{Yoon2017g, Shaaban-etal-2021b} and one- and two-dimensional particle-in-cell (PIC) simulations \citep{Micera-etal-2020,Riquelme2018}.
These investigations are motivated not only by the solar wind observations, which show that both the electrons and protons may satisfy conditions for the excitation of FIs \citep{Hellinger-et-al-2006, Stverak-etal-2008}, but also by the different (temporal and spatial) scales at which the EF and PF instabilities are expected to develop \citep{Michno-etal-2014, Maneva-etal-2016, Micera-etal-2020, Shaaban-etal-2021b}. 
When both the electron and proton populations are anisotropic with $A_{e,p} < 1$, the combined linear spectrum is dominated by the a-EFI, which has the highest growth rate \citep{Maneva-etal-2016}. 
Thresholds of p-PFI are also influenced by anisotropic electrons, suggesting that this instability may also contribute to proton anisotropy constraint, thus explaining solar wind observations \citep{Michno-etal-2014,Shaaban-etal-2017}. 
However, it remains to be clarified how the thresholds of the a-PFI are affected, and how the regimes of dominance of each of these two branches change in this case. 
The parametric regimes dominated by the p-PFI developing along ${\bf B}_0$ have been modeled by a set of 1D PIC (particle-in-cell) simulations \citep{Micera-etal-2020}. In the presence of electrons with $A_e < 1$ the onset of this instability occurs earlier with a higher growth rate, leading to enhanced wave fluctuations, which reduce the proton temperature anisotropy to a stable, nearly isotropic state. 
At oblique angles with respect to ${\bf B}_0$, the 1D spectra of magnetic wave energy confirm linear theory predictions, revealing that the a-EFI develops faster than p-PFI, but also saturates very fast. 
This evolution is followed by the more pronounced peaks of PFIs: p-PFI, detectable only in parallel direction and at small angles, and then the other branch of a-PFI, which grows later in time but reaches comparable or even higher levels of magnetic wave energy at highly oblique angles of propagation. The same succession of these PFI branches has also been confirmed in hybrid simulations of ideal cases with isotropic ($A_e= 1$) electrons \citep{Hellinger-Matsumoto-2001}.

In this paper we present results from 2D PIC simulations of the interplay of firehose instabilities, as conditioned by the anisotropy of electrons and protons with $A_e < 1$ and $A_p <1$, respectively.
In contrast to previous studies, here we focus on different regimes of PFIs, which ensure favorable conditions for a first ignition and development of a-PFI, before the growth of p-PFI branch (see section \ref{sim}).
A comparative analysis is provided for the PFIs, contrasting the case of isotropic and anisotropic electrons\footnote{We draw attention to the fact that in-situ measurements capture only those quasi-stable plasma states of low temporal resolution, which accumulate at small deviations from isotropy \citep{Stverak-etal-2008, Bale-etal-2009}, below the thresholds of instabilities discussed here.} exhibiting $A_e < 1$. 
We first obtain new insights from linear theory (in section \ref{theory}) by describing the effects, not only on the growth rates of EFIs and PFIs, but also on the parametric conditions of the more robust PFIs, showing when their spectra are dominated either by the p-PFI or by the a-PFI. 
These results allowed us to make the right choices for the setup of numerical simulations discussed in section~\ref{sim}. These 2D-PIC simulations help us to understand not only the interplay of EFI and PFI modes, which are initiated at very different time scales, but also to differentiate between a-PFI and p-PFI modes, whose competition, we will see, becomes even stronger with increasing electron anisotropy.
The last section (section \ref{conclusions}) summarizes the main conclusions of our analysis, as well as their implications for a realistic modeling and interpretation of the properties of space plasmas.

\section{New insights from theory}\label{theory}

This section presents new linear properties of firehose instabilities, when both electron and proton populations are anisotropic, with $A_e <1$ and $A_p <1$, respectively. 
We have resolved the general dispersion relation for bi-Maxwellian plasma, using the numerical solver DIS-K, capable to resolve the full spectrum of wave instabilities in the whole wave-vector (${\bf k}$) space \cite{Lopez2021}. 
We thus reveal new changes in the EFI and PFI spectra, under the mutual effects of anisotropic electrons and protons.
These results complement previous linear analyzes \citep{Michno-etal-2014, Maneva-etal-2016}, and also help to understand the long-run evolution from numerical simulations in section \ref{sim}.

\begin{figure}[t!] 
\centering
\includegraphics[width=0.48\textwidth]{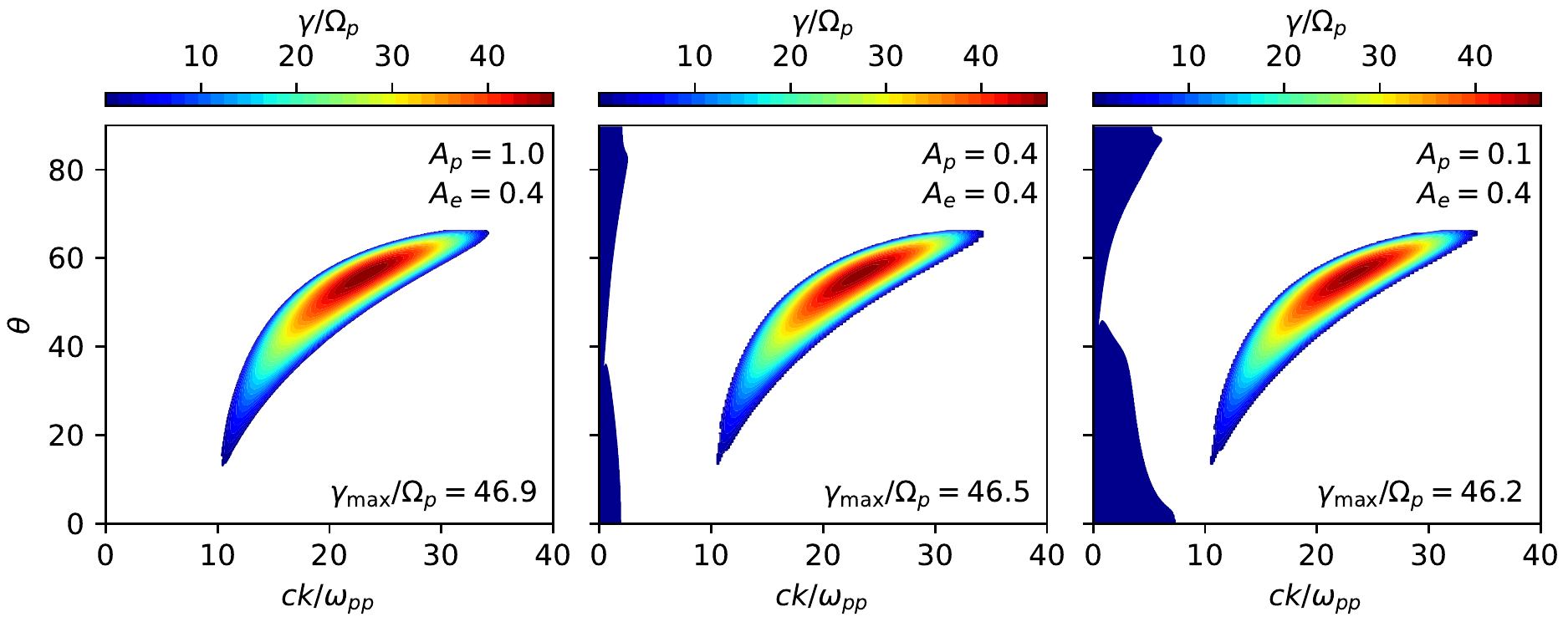}
\caption{Growth rates (color coded) in the $k$--$\theta$ plane, derived for $A_e=0.4$, $\beta_{e,\parallel} = \beta_{p,\parallel} = 2.5$, and different proton anisotropies, $A_p = 1.0$ (left), $0.4$ (middle) and $0.1$ (right). All cases are dominated by the a-EFI at oblique angles, contrasting to p-EFI at small angles (only minor signature in the right panel) and similar branches of PFI with  very low growth rates (blue) at lower wave-numbers (if $A_p < 1$).} \label{efhi-p-an}
\end{figure}

\subsection{Electron firehose instabilities (EFIs)}

The main parameters in this analysis are the parallel plasma beta $\beta_{j,\parallel}=8\pi n_jk_BT_{j,\parallel}/B_0^2$, i.e., component parallel to the background magnetic field, and the temperature anisotropy $A_j= T_{j,\perp} / T_{j,\parallel}$, where $n_j$ is the total density of species $j$, $k_B$ is the Boltzmann constant, and $B_0$ the background magnetic field.  
For the solar wind plasma conditions, when electrons and protons show comparable values of these parameters, linear theory predicts spectra dominated by the aperiodic branch of the a-EFI. This is found at oblique propagation directions and with growth rates higher than those of the periodic p-EFI, and of course, much higher than the growth rates of the PFIs. 
We are thus fully motivated to begin our analysis with EFIs, triggered by anisotropic electrons with $A_e <1$.

Figure~\ref{efhi-p-an} displays growth rates of EFIs (color coded) derived for a parameterization similar to previous studies \citep{Maneva-etal-2016, Micera-etal-2020}, i.e., $A_e = 0.4$, $\beta_{e,\parallel}=\beta_{p,\parallel} = 2.5$. Here the wavenumber is normalized to the proton inertial length, $c/\omega_{pp}$, and the growth rates are normalized to the proton gyro-frequency, $\Omega_p$, with $\omega_{pj}=(4\pi n_je^2/m_j)^{1/2}$, and $\Omega_j=eB_0(m_jc)$, the plasma and gyro-frequency, respectively.
In Figure~\ref{efhi-p-an} we compare the ideal case with isotropic protons, i.e., $A_p = 1.0$ (left), with two situations when protons are anisotropic $A_p= 0.4$ (middle) and 0.1 (right).
Found at highly oblique propagation angles $\theta$, the a-EFI branch dominates the spectra, having maximum growth rates (given in each panel) much higher than all the other branches. 
The electron parameters are below the threshold of the p-EFI branch (at low angles $\theta$), which is not observed in this case (left panel). 
However, for anisotropic protons with $A_p <1$ (middle and right panels) both the p-PFI and a-PFI branches are obtained, and both have growth rates much lower than a-EFI.
There is no noticeable influence of proton anisotropy ($A_p < 1$) on the fastest growing a-EFI, as the maximum growth rate is reduced by only about 1~\%, and the propagation angle remains almost unchanged. 

%
\begin{figure}[t!] 
\centering
\includegraphics[width=0.47\textwidth]{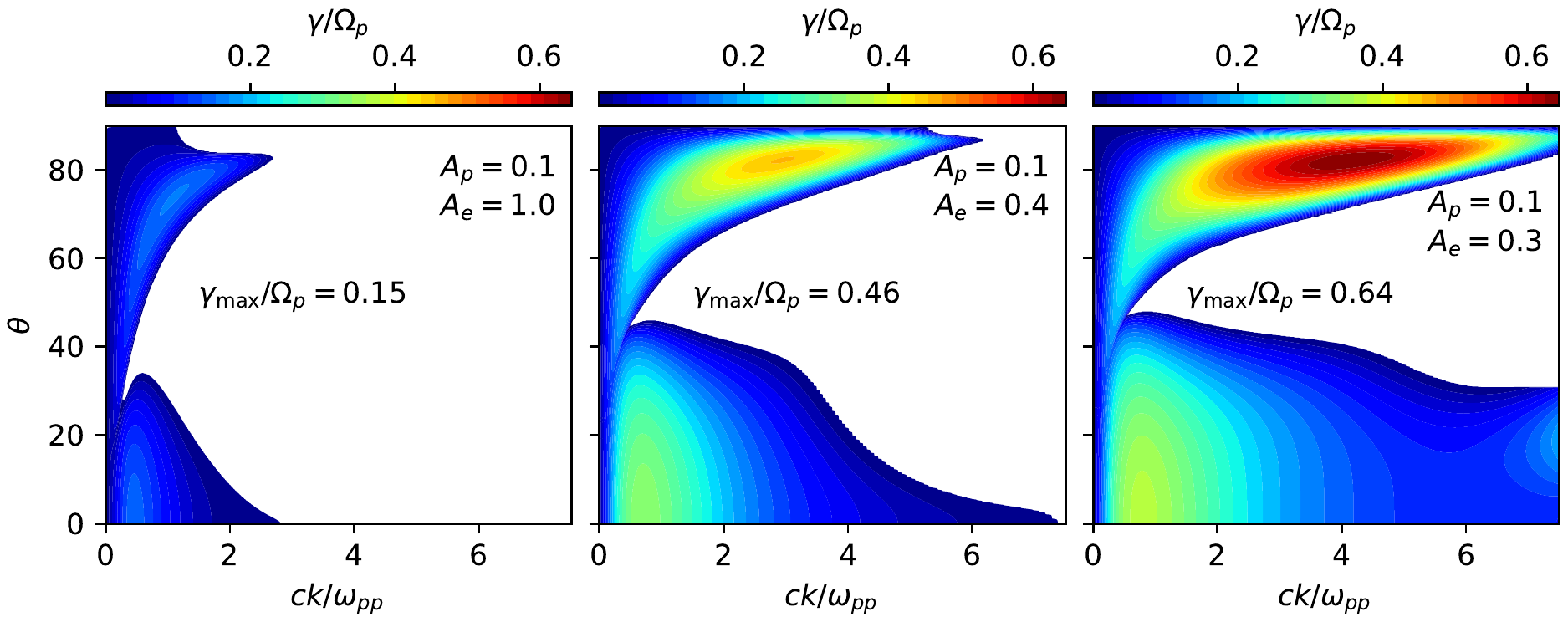}
\includegraphics[width=0.47\textwidth]{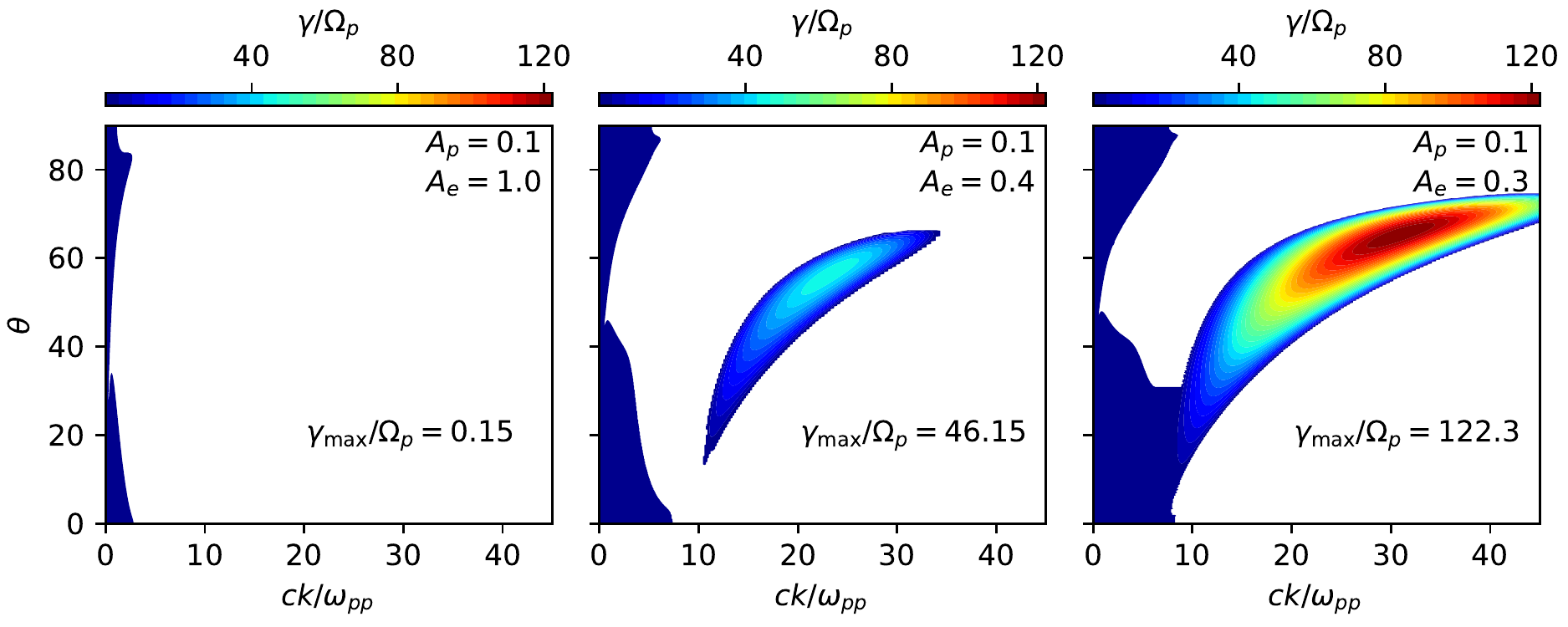}
\caption{Growth rates (color coded) in the $k$--$\theta$ plane, derived for $A_p=0.1$, $\beta_{e,\parallel} = \beta_{p,\parallel} = 2.5$, and different electron anisotropies, $A_e=1.0$ (left), $0.4$ (middle) and $0.3$ (right). The top panels focus on PFI branches at lower wavenumbers, while the bottom panels show the full spectra.} \label{pfhi-e-an} 
\end{figure}

\subsection{Proton firehose instabilities (PFIs)}

%
\begin{figure*}[t!] 
\centering
\includegraphics[width=0.3\textwidth]{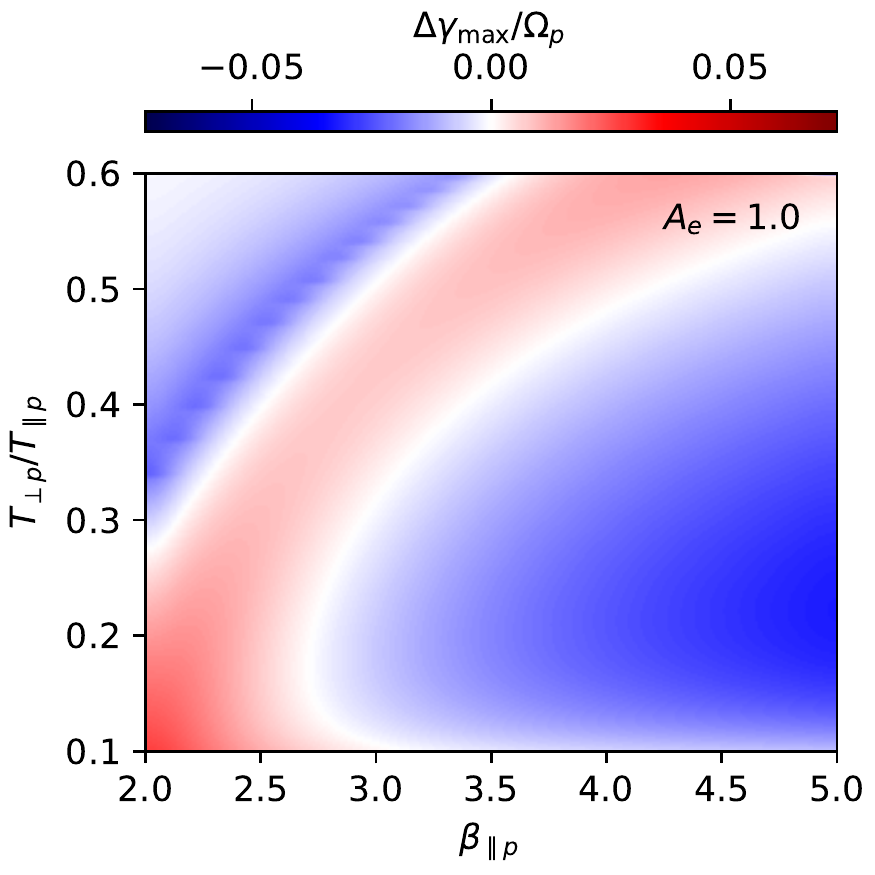}
\includegraphics[width=0.3\textwidth]{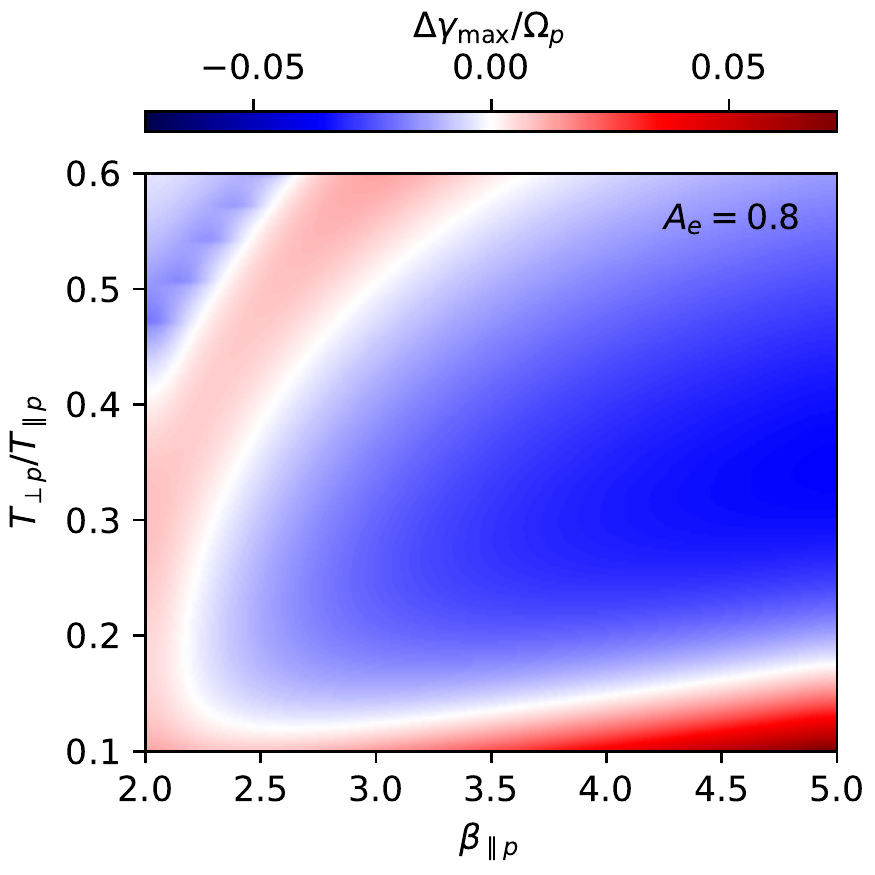}
\includegraphics[width=0.3\textwidth]{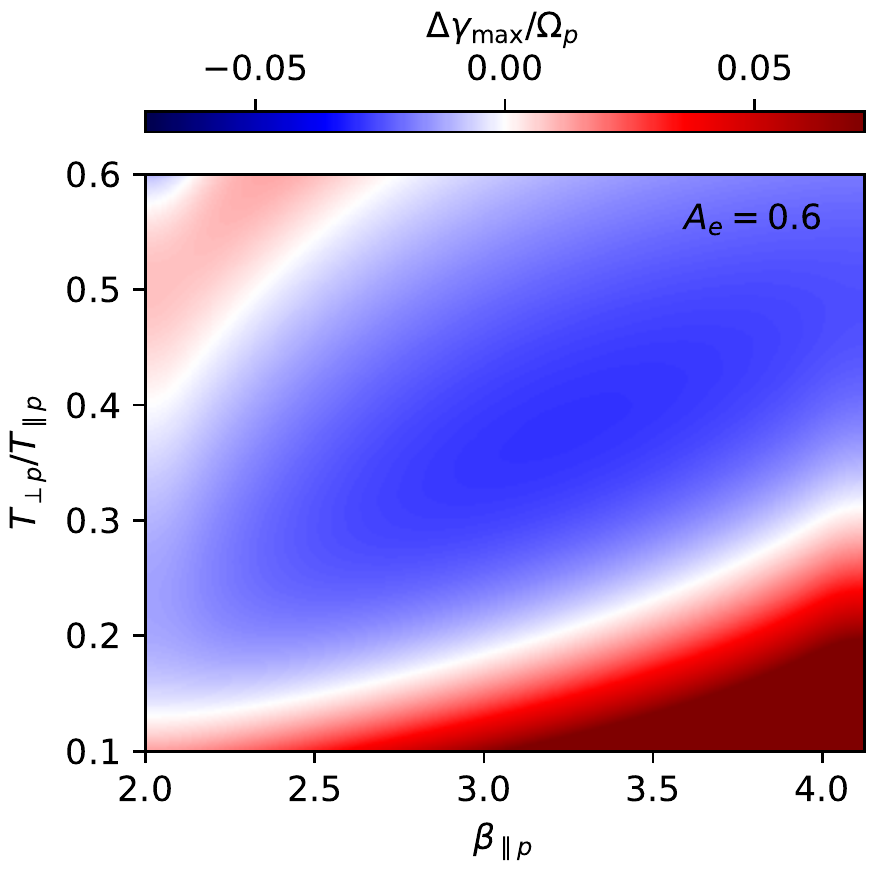}
\caption{Influence of electron anisotropy on the maximum growth rate difference between a-PFI and p-PFI, $\Delta \gamma_\text{max} = \gamma_\text{max} (\text{a-PFI})-\gamma_\text{max}(\text{p-PFI})$, for $A_e=1$ (left), $A_e=0.8$ (center) and $A_e=0.6$ (right). Reddish shades indicate a dominance of a-PFI, while blue shades are regimes favorable to p-PFI.} \label{gmax-r} 
\end{figure*}

Now, let us see what happens to the spectra of PFIs, when electrons are  anisotropic with $A_e <1$. 
A competition between the aperiodic (a-PFI) and periodic (p-PFI) branches is expected, dominated by the latter \citep{Hellinger-Matsumoto-2000,   Micera-etal-2020}. 
For isotropic electrons ($A_e =1$), linear theory predicts extended regimes of a dominant p-PFI branch, with growth rates higher than those of a-PFI \citep{Hellinger-Matsumoto-2000}, confirmed via one- and two-dimensional hybrid PIC simulations \citep{Hellinger-Matsumoto-2001}. 
For certain conditions, in the presence of anisotropic electrons with $A_e<1$, 1D PIC simulations have also shown the dominance of this mode \citep{Micera-etal-2020}.

These two branches can be distinguished in Figure~\ref{pfhi-e-an}, with details visible in the top panels: the a-PFI with a maximum growth rate at highly oblique angles, and the p-PFI propagating at small angles and with a maximum growth rate along the magnetic field direction.
Unstable spectra are obtained for $A_p = 0.1$ and different electron anisotropies $A_e = 1.0$ (left), $0.4$ (middle) and $0.3$ (right), which show a significant increase of the (maximum) growth rates, of both the p-PFI and a-PFI branches, under the influence of anisotropic electrons with $A_e <1$. 
For this set of parameters, both a-PFI and p-PFI have comparable growth rates when electrons are isotropic (with the first slightly dominating), with $\gamma_\text{max}/\Omega_p=0.15$ and $\gamma_\text{max}/\Omega_p=0.14$, respectively. 
The growth rates of both branches are enhanced as the electron (initial) anisotropy increases, but the peaking (maximum) growth rates of a-PFI become markedly higher than those of p-PFI.
Thus, for a-PFI we find $\gamma_\text{max}/\Omega_p = 0.15$ for isotropic electrons, and then $\gamma_\text{max}/\Omega_p=0.46$ and $\gamma_\text{max}/\Omega_p=0.64$ for $A_e=0.4$ and $A_e=0.3$, respectively. 
There is also an effect on the range of unstable wave-numbers, which increases significantly for both a-PFI and p-PFI. 
In the last case ($A_e = 0.3$), the presence of p-EFI also becomes visible at large wave-numbers, which are covered in more detail in the bottom panels. For anisotropic electrons, the spectra are clearly dominated by the a-EFI (as in Figure~\ref{efhi-p-an}), with growth rates highly increasing with increasing the electron anisotropy.

It seems that anisotropic electrons have a more stimulating effect on the a-PFI.
This is indeed what we found not only for these cases but also for the unstable solutions derived for an extended range of proton parameters. 
Thus, Figure~\ref{gmax-r} displays the difference in maximum growth rates (normalized by the proton gyrofrequency $\Omega_p$) $\Delta \gamma_{\rm max} = \gamma_{\rm max}(\text{a-PFI}) - \gamma_{\rm max}(\text{p-PFI})$, derived as a function of proton temperature anisotropy $T_{p,\perp}/T_{p,\parallel}$ and proton parallel beta $\beta_{p,\parallel}$ for $A_e=1$ (left), $A_e=0.8$ (center) and $A_e=0.6$ (right), using the same color scale for all three cases. 
The plot in the left panel (isotropic electrons) is obtained for the same set of parameters used in Plate~3 in \cite{Hellinger-Matsumoto-2000}. 
In this case, a-PFI dominates, i.e., $\Delta\gamma_{\rm max} > 0$, in a relatively narrow sub-range of parameters displayed with reddish shades. 
As the electron anisotropy increases, the ranges of dominance change, showing that a-PFI can be faster than p-PFI even for larger $\beta_{p,\parallel}$ away from the instability thresholds. 
The reddish area is slightly narrowed and moves toward threshold conditions to lower values of $\beta_{p,\parallel} < 2$. 
If the electrons are sufficiently anisotropic (third panel), the difference becomes even more significant in favor of a-PFIs, which dominate not only for large values of betas but also for moderate values of anisotropy.
Thus, the dominance of a-PFI (reddish area) extends to all values of $\beta_{p,\parallel} > 1$ considered (already in the second panel) if deviation from proton isotropy is sufficiently high, i.e., $A_p \leqslant 0.1$ (this limit increases for higher values of $\beta_{p,\parallel}$).
Of interest here are these regimes which, unlike other regimes discussed before \citep{Micera-etal-2020}, should be dominated by a-PFI instability, see linear solutions in Figures~\ref{efhi-p-an} and \ref{pfhi-e-an}. In the next section, we use numerical simulations to verify these predictions. 
\begin{table}[t!]
\caption{Plasma parameters of the analyzed cases}
\centering
\begin{tabular}{l l c c c c}
\hline\hline
 Case & Configuration& $\beta_{p, \parallel}$&$\beta_{e, \parallel}$&$A_{p}$&$A_e$ \\ [0.5ex]
\hline
 1 & PFI & 2.5 & 2.5 & 0.1 & 1.0  \\
 2 & PFI + EFI  & 2.5 & 2.5 & 0.1 & 0.4 \\ 
 3 & PFI + EFI  & 2.5 & 2.5 & 0.1 & 0.3 \\[1ex] 
\hline
\end{tabular} \label{t1}
\end{table}

Particularly important are the plasma conditions around the following values of the proton parameters, i.e., $\beta_{p,\parallel} = 2.5$ and $A_p = 0.1$ (see also Table~\ref{t1}), which remain relevant for the competition of PFI branches near the threshold conditions (see the reddish shades in all the panels in Figure~\ref{gmax-r}), and also ensure a reasonable duration of the numerical simulations.

\section{Results from PIC simulations}\label{sim}

We consider such cases in the following PIC simulations, in order to describe the time evolution of the complex spectra generated by the interplay of EFIs and PFIs, when both populations of electrons (subscript $e$) and protons (subscript $p$) are anisotropic, respectively, with $A_e <1$ and $A_p <1$ (cases 2 and 3 in Table~\ref{t1}).

\subsection{Plasma parameterization and simulations setup}\label{param}

The plasma parameterization considered in our analysis is summarized in Table~\ref{t1}. We show the values of parallel plasma beta parameters ($\beta_{e,\parallel}$, $\beta_{p,\parallel}$) and anisotropies ($A_e$, $A_p$) of electrons and protons, which are the main parameters needed in the dispersion and stability analysis.
For comparison, included are the interplay (EFI + PFI) conditions, when both the electron and proton populations are anisotropic, i.e., $A_{e,p} < 1$ in cases 2 and 3, and also the individual case 1 of $A_e = 1$ and $A_p < 1$, when only the PFI branches are excited. Case 3 does not differ much from case 2, being mainly intended to test the consistency of the results in PIC simulations.

In the present analysis, the parameterization of the plasma populations is highly conditioned by the available resources for our numerical simulations.
This restricts us, especially in choosing the proton parameters, to values large enough for $\beta_{p,\parallel} >1$, and sufficiently small for $A_p < 1$ (meaning large enough deviations from isotropy), in order to increase the growth rate of the instabilities simulated and consequently reduce the time required for their saturation to be observed on both electron and proton scales. For such situations, a-PFI is expected to be more effective and develop faster than p-PFI, see Figure \ref{gmax-r}. 
Moreover, to capture the dynamic of both protons and electrons, we use the semi-implicit PIC code, iPic3D \citep{Markidis2010}, which allows us to resolve the multiple spatial and temporal scales by removing the most rigid stability constraints typical of explicit PIC codes~\citep{Lapenta2016}. 
Since in our simulations we want to simultaneously discern the physics at electron scales and conduct analyses covering the longer timescales characteristic of proton instabilities, simulations using the real mass ratio are beyond our reach in terms of computational capabilities. Thus, we choose a reduced mass ratio $m_p/m_e=100$, for which the spatial and temporal scales of both PFI and EFI are distinguishable enough to study both behaviors but close enough to do it in a reasonable time.
In our simulations, we consider $\omega_{pe} / \Omega_{ce} = 20$. Our simulation box is of size $L_x=L_y=45\,c/\omega_{pp}$ and has been discretized with cells of size $\Delta x=\Delta y=0.074\,c/\omega_{pp}$. Particles were pushed for 209,000 (case 1) and 178,000 (cases 2 and 3) iterations with a time step of $\Delta t=0.2/\Omega_p$. We use 1024 particles per cell per species. The remaining initial values of the main parameters are given in Table~\ref{t1}.

Using a reduced mass ratio $m_p/m_e=100$ in simulations is supported by the results obtained from linear theory in Appendix~\ref{ApA}. 
The effect of the reduced mass ratio on the growth rates can be estimated if we compare Figure~\ref{pfhi-e-an} with Figure~\ref{disp_mpme} from Appendix~\ref{ApA}. 
Figure~\ref{disp_mpme} uses a  $k_x$--$k_y$ plane to display the growth rates derived for the same cases 1, 2 and 3. 
The left panels focus on the PFI at smaller wavenumbers, while the right panels show the complete spectra of unstable modes. 
As expected, a lower mass ratio reduces the differences between the maximum growth rates of PFI and EFI, and between their specific ranges of unstable wavenumbers. 
However, the effect of anisotropic electrons on the PFI branches does not change much, showing the same enhancement of the growth rates and the unstable range of wavenumbers. In all three cases of Table~\ref{t1}, for $A_p=0.1$, a-PFI dominates over p-PFI, as can also be see in Figures~\ref{disp_mpme} and~\ref{gmax} from Appendix~\ref{ApA}.
Figure \ref{gmax} shows the maximum growth rate difference between p-PFI and a-PFI, as in Figure \ref{gmax-r}, but this time for a reduced mass ratio, i.e., $m_p/m_e=100$. 
The regimes of dominance remain almost unchanged, and there are only subtle differences in the case of isotropic electrons. Thus, around $T_{\perp p}/T_{\parallel p} \approx 0.1$ the a-PFI dominates for all betas, in contrast to the real mass ratio case, where p-PFI is dominant only for $\beta_{\parallel p}>3.0$.

\begin{figure}[t!] 
\centering
\includegraphics[width=0.48\textwidth]{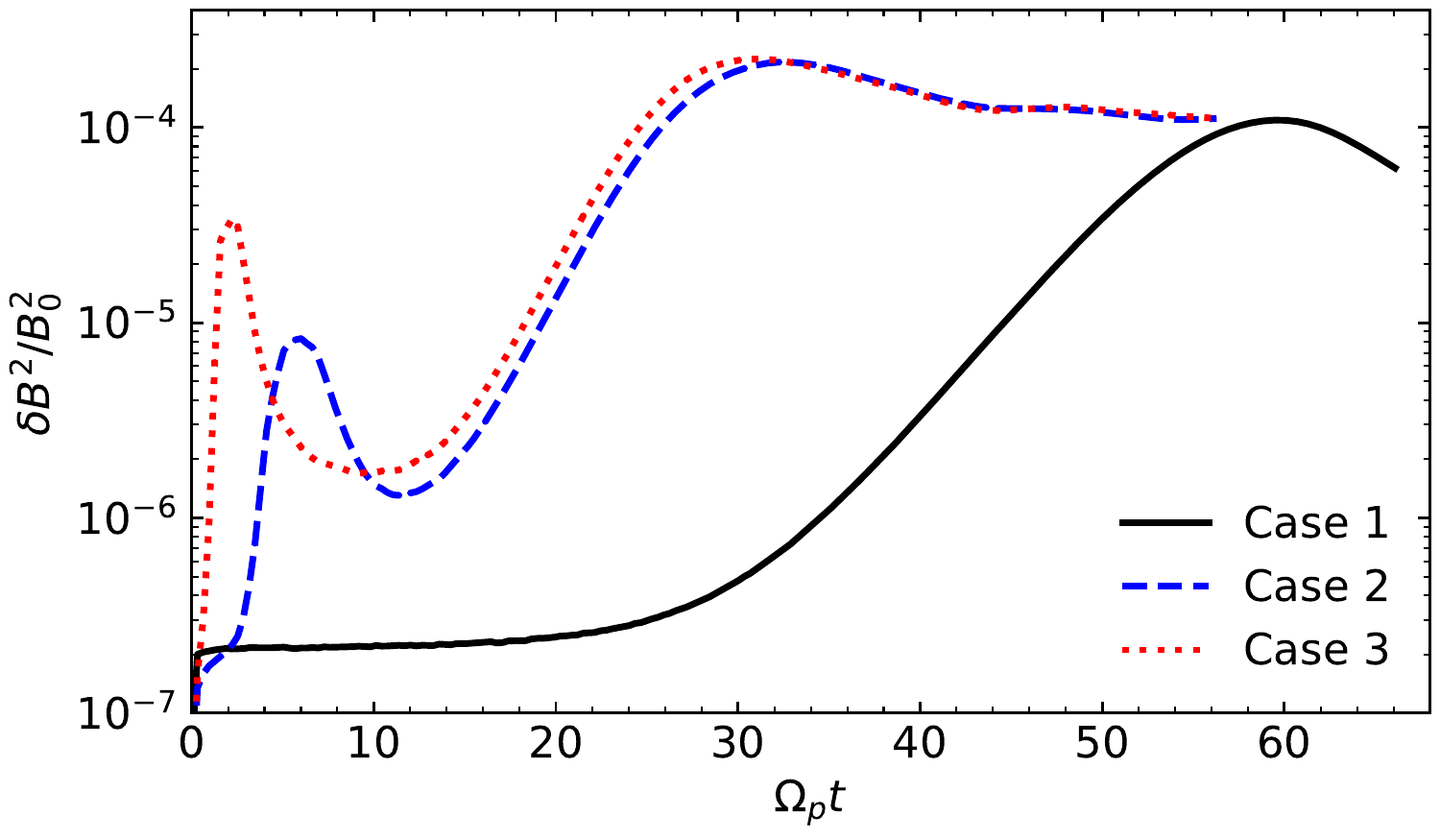}
\caption{Time evolution of the total magnetic wave energy corresponding to the simulation sets in Table~\ref{t1}.}\label{energy}
\end{figure}

\subsection{PFIs $+$ EFIs ($A_p < 1$ and $A_e < 1$)}

\begin{figure}[t!] 
\centering
\includegraphics[width=0.48\textwidth]{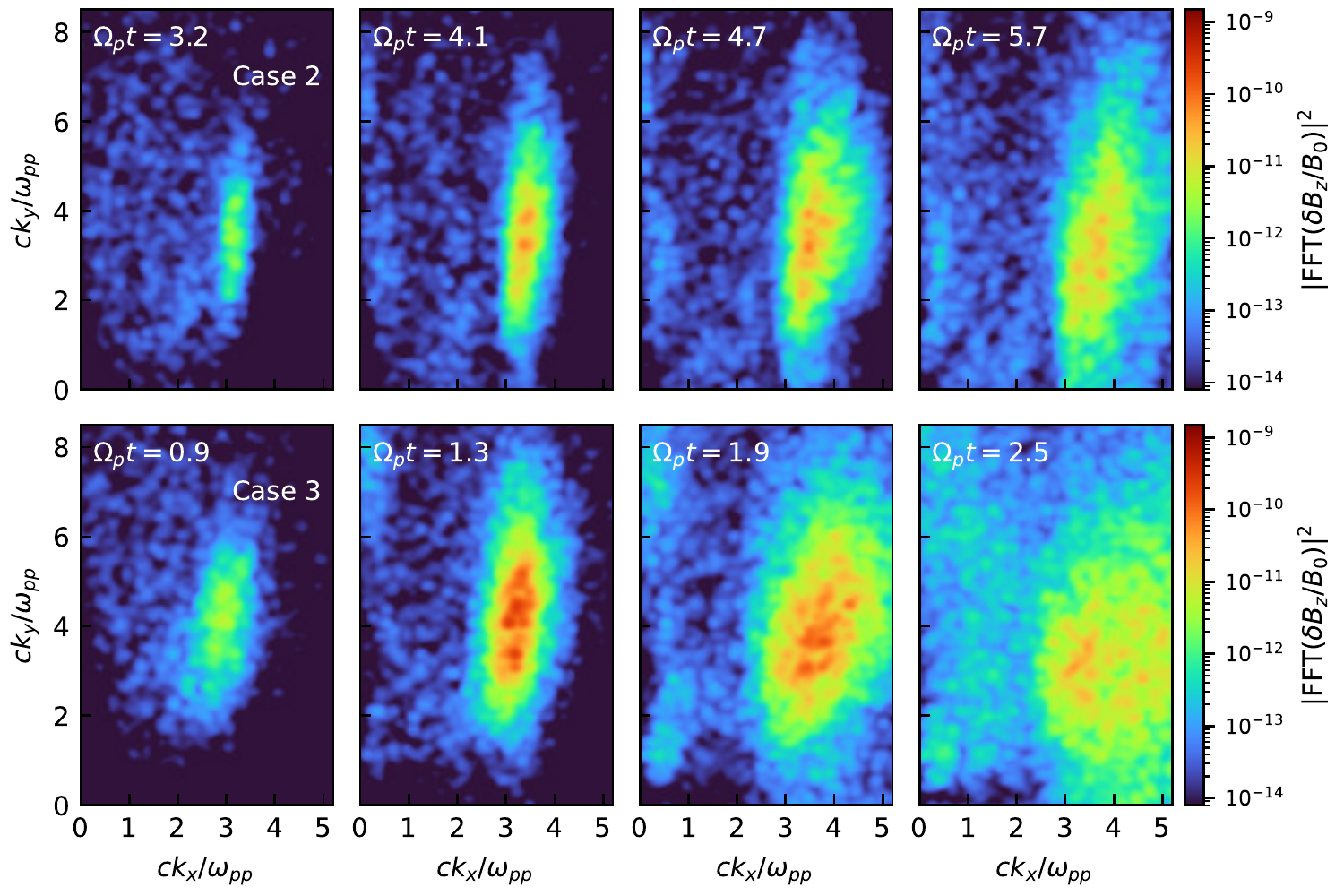}
\caption{Power spectra of the transverse magnetic fluctuations, $|{\rm FFT}(\delta B_z/B_0)|^2$, at electron scales for cases 2 and 3 in Table~\ref{t1}.}\label{spectra-EFI}
\end{figure}

In this section, we analyze the results obtained from the PIC simulations for the three cases in Table~\ref{t1}. 
Figure~\ref{energy} displays the time evolution of the total magnetic wave energy, $\int dxdy\,\delta B^2/B_0^2$, obtained from the simulations for all these cases. 
As we might expect, the peak corresponding to the a-EFI branch arises on very short time scales for the cases when electrons are anisotropic (cases~2 and 3 plotted with blue-dashed and red-dotted lines). 
The nature of this a-EFI, developing only at oblique angles, is confirmed by the power spectra in Figure~\ref{spectra-EFI}.
The panels in this figure show the power spectra of the transverse magnetic fluctuations, $|{\rm FFT}(\delta B_z/B_0)|^2$ (color coded), as a function of $k_x$ and $k_y$, for case 2 (top panels) and case 3 (bottom panels), at different times before reaching the saturation.  The peak intensity, as well as the range of unstable wavenumbers, are higher in case 3, consistent with the linear prediction in Appendix~\ref{ApA}.
The details in the spectra show a conversion in time to lower propagation angles, where the spectrum combines with the p-EFI branch. 
As already mentioned, the saturation of the a-EFI shown in Figure~\ref{energy} is quite fast, and occurs well before any reaction from the protons. After saturation, the wave energy drops abruptly to much lower (one order of magnitude lower) values.

Figure~\ref{energy} also shows that the magnetic energy of the fluctuations, after dropping to a local minimum following the saturation of the EFI, starts to grow again exponentially, reaching a value of one order of magnitude larger than the EFI. The growth is due to the PFI, whose onset appears to occur at earlier times in the presence of anisotropic electrons. By comparing the growth rate of the PFI in the presence of anisotropic electrons and protons with the growth rate of the same instability with only anisotropic protons (black curve in Figure~\ref{energy}), we can observe that the growth of the PFI is enhanced by the electron anisotropy, thus confirming the prediction of linear theory (see Figures~\ref{pfhi-e-an} and \ref{disp_mpme}). Finally, regarding the evolution of the PFI, there is not much difference between cases 2 and 3, with the former reaching a slightly higher energy peak.

\begin{figure}[t!] 
\centering
\includegraphics[width=0.48\textwidth]{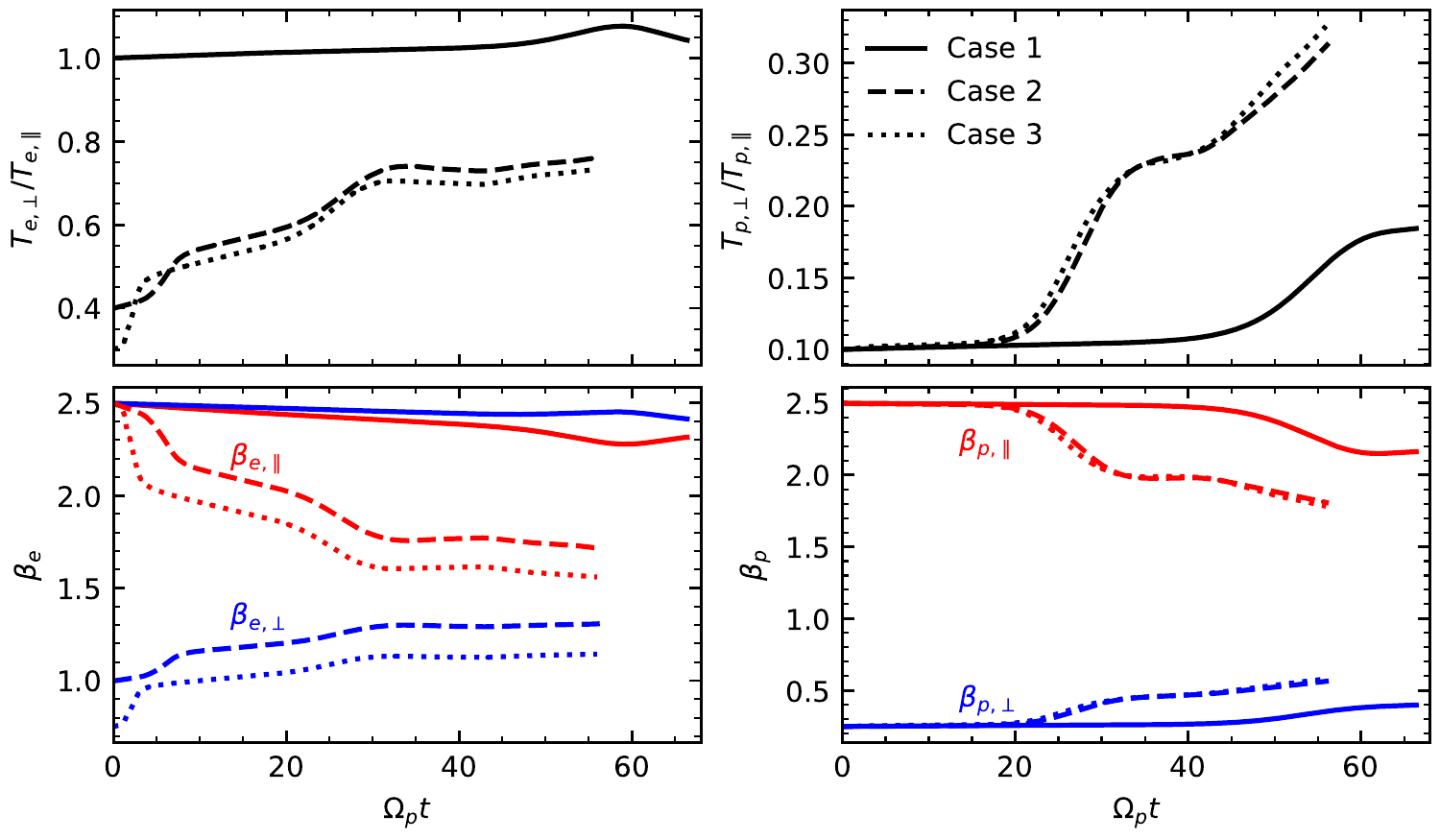}
\caption{Time evolution of the temperature anisotropy of electrons (top left) and  protons (top right), and the corresponding parallel and perpendicular components of the electron plasma beta (bottom left) and the proton plasma beta (bottom right), for cases 1, 2 and 3 in Table~\ref{t1}.}\label{betas}
\end{figure}

The effect of the enhanced fluctuations on the anisotropy of electrons and protons, are shown in Figure~\ref{betas}, left and right panels, respectively. 
The three cases are shown with solid (case 1), dashed (case2), and dotted (case 3) lines. As expected, electrons react quickly to the instability, showing a rapid reduction of the anisotropy, which leads to an increase of the perpendicular plasma beta and a reduction of the parallel beta. 
This effect is more pronounced in case 3. Still, after the a-EFI is saturated, the behavior of the anisotropy and betas follows a similar trend in cases 2 and 3. 
Electron relaxation is not smooth, but occurs in steps, or distinct phases, and continues after the saturation of the EFI. This phenomenon has been already observed in the quasi-linear theory of the interplay of anisotropic electrons and protons \citep{Shaaban-etal-2021b}. 
The third more pronounced phase of relaxation, i.e., after $t >20\,\Omega_p^{-1}$, seems to be triggered by the growing PFI fluctuations, see Figure~\ref{energy}. 
There is also a fourth phase, when the electron anisotropy stagnates or slightly increases, and then decreases again. 
This non-monotonic behavior may be caused by a selective interaction of electrons with a-PFI and p-PFI. The latter, which develops later, is right-handed (RH) polarized and can possibly be responsible for a perpendicular resonant heating of electrons. 
Moreover, on the time scales corresponding to PFI  fluctuations, the temperature anisotropy of protons increases (e.g. evolves toward isotropization), as can be seen in the top-right panel of Figure~\ref{betas}. This is mainly due to the decrease of $T_{p,\parallel}$, and the corresponding $\beta_{p,\parallel}$, as shown in the bottom-right panel. The evolution of the proton anisotropy is markedly accelerated in the presence of anisotropic electrons (cases 2 and 3), and leads to lower anisotropies of protons, i.e., a more efficient isotropization.

\begin{figure}[t!] 
\centering
\includegraphics[width=0.48\textwidth]{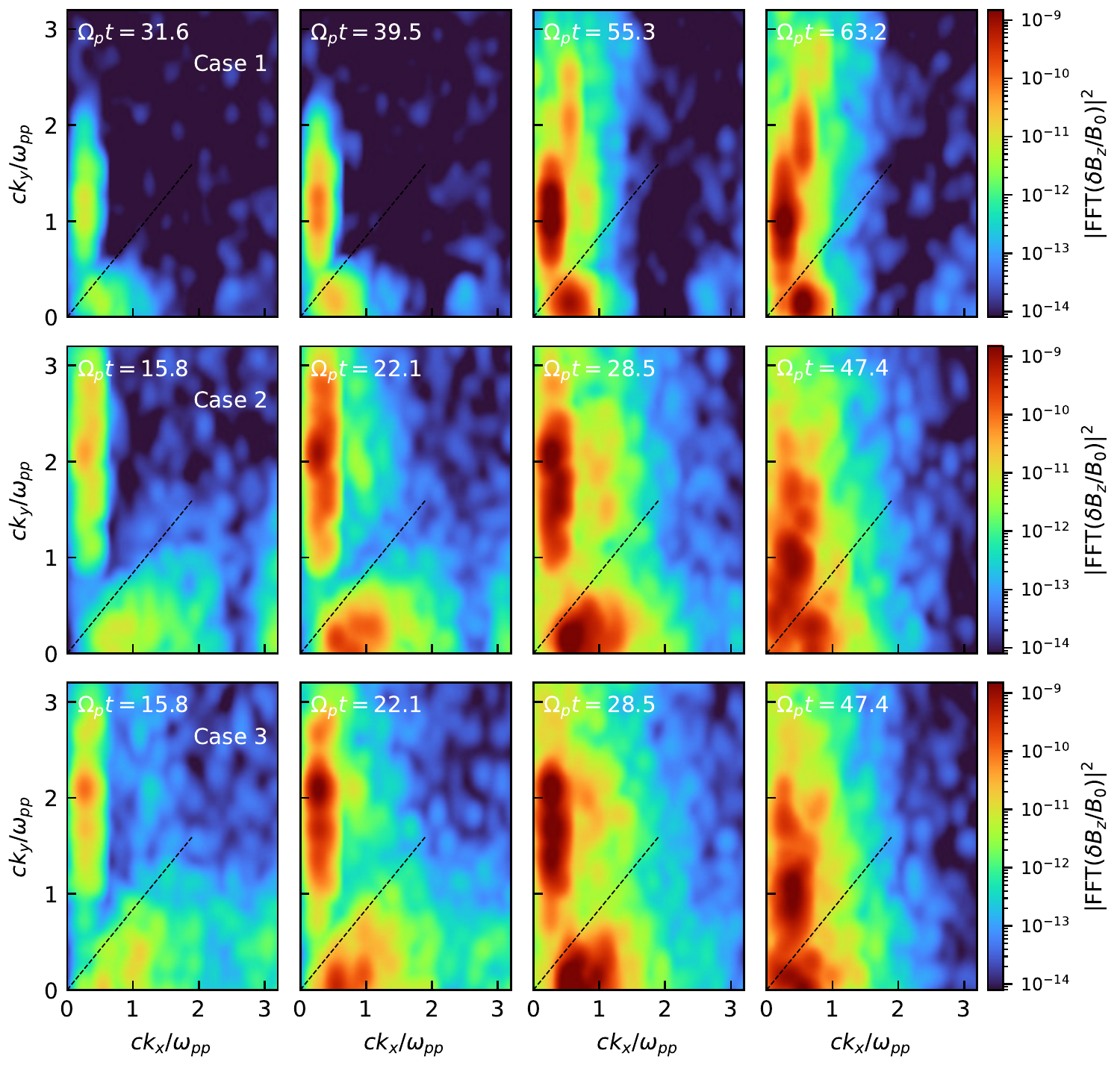}
\caption{Power spectra of the transverse magnetic fluctuations, $|{\rm FFT}(\delta B_z/B_0)|^2$, for the PFI branches in cases 1 (top), 2 (middle) and 3 (bottom) in Table~\ref{t1}. Dashed line indicates the angle $\theta=40^\circ$.}\label{spectra-PFI}
\end{figure}

Now let us unveil more details about the nature of PFIs modes during their initiation and development in our simulations. 
Similar to Figure~\ref{spectra-EFI} for EFI, Figure~\ref{spectra-PFI} displays the power spectra of PFI branches. It shows fluctuations of the a-PFI at highly oblique angles, i.e., for $k_y > k_x$, and of the p-PFI at lower angles and along the magnetic field direction, i.e., for $k_y < k_x$. 
These spectra are computed for the three cases in Table~\ref{t1}, corresponding to the top, middle, and bottom panels, respectively, and at different times (increasing from left to right). 
The dashed line in the panels represents an attempt to separate the contribution of the power spectra of the a-PFI ($k_y>k_x$) from the contribution of the p-PFI ($k_y<k_x$). The lines lies at an angle of $40^\circ$ with respect to $k_x$.
This limit may need adjustments, especially for the late times (e.g., in the last column), but in the linear phase of the instability it agrees quite well with the linear spectra in Figure~\ref{disp_mpme} in Appendix~\ref{ApA}. 
The a-PFI branch dominates the initial phases, but its relative importance is reduced over time. 
For example, at the time corresponding to the third column, the power in the fluctuations is comparable for both a-PFI and p-PFI.

\begin{figure}[t!] 
\centering
\includegraphics[width=0.48\textwidth]{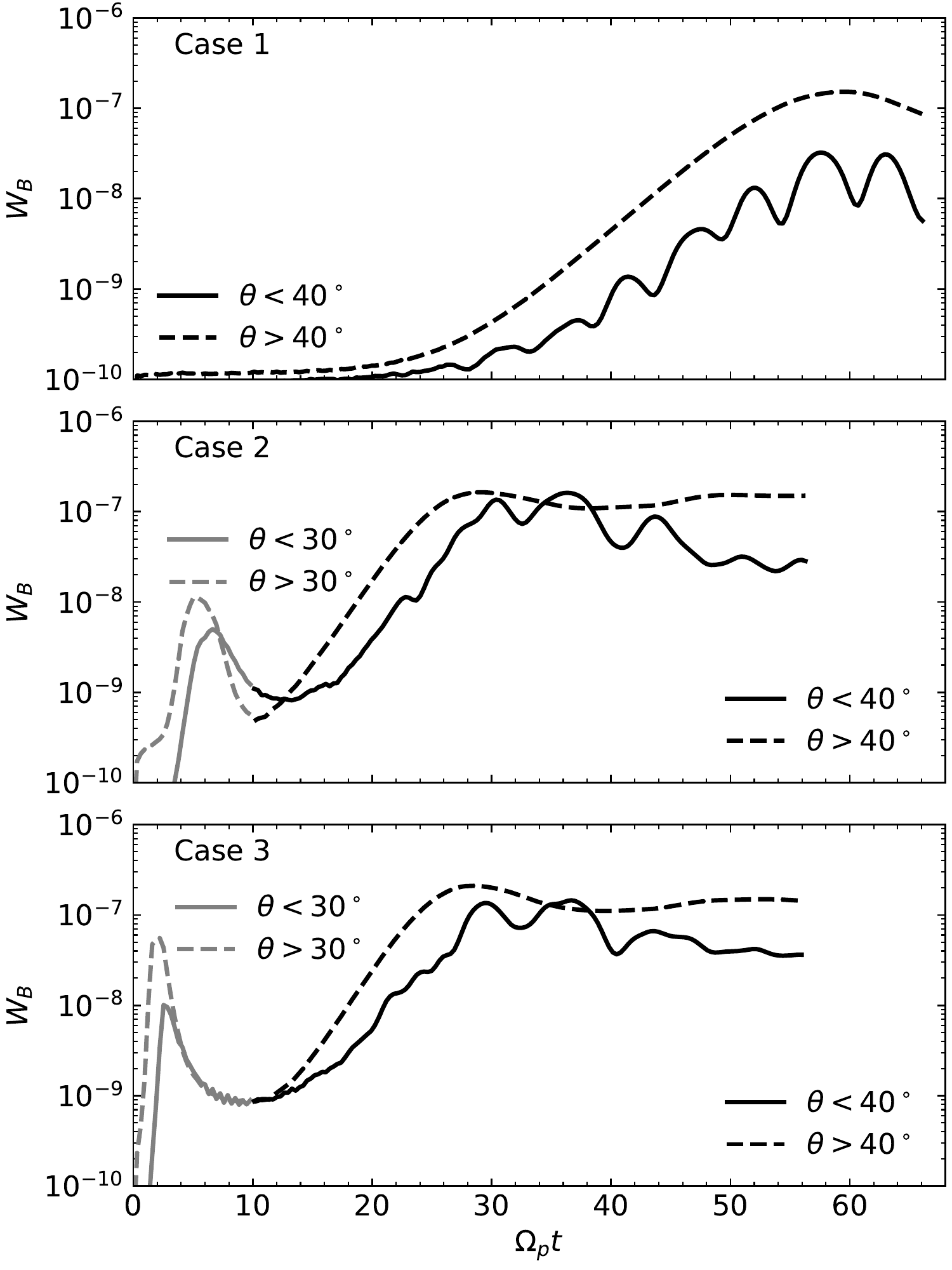}
\caption{Temporal evolution of the integrated magnetic energy spectra at all angles above and bellow of the angular delimitation from Figure~\ref{spectra-PFI}, for quasi-parallel ($\theta<40^\circ$) and oblique modes ($\theta>40^\circ$), for cases 1-3 in table~\ref{t1}. For cases 2 and 3 we have used $\theta=30^\circ$ as a limit in the initial stage of the simulation (in gray), $\Omega_{p}t<10$, where a-EFI is operative, see Fig.~\ref{spectra-EFI}.}\label{spectra_energy2}
\end{figure}

We confirmed these qualitative estimates by integrating the power spectra at all angles below and above the angular delimitation between aperiodic (highly-oblique angles) and the periodic modes at lower angles, as $W_B=\int d\mathbf{k}\,|\text{FFT}(\delta B_z/B_0)|^2$ in Figure~\ref{spectra_energy2}. The distinction between the modes is is roughly given by the black line in Figure~\ref{spectra-PFI}, which allows us to differentiate between p-PFI for $\theta < 40^{\rm o}$ and a-PFI for $\theta > 40^{\rm o}$ and for earlier times, as in Figure~\ref{spectra-EFI}, p-EFI for $\theta < 30^{\rm o}$ and a-EFI for $\theta > 30^{\rm o}$, shown in gray. These are plotted with solid and dashed lines, respectively  (see legends), and are displayed for all three cases in Table~\ref{t1}. It becomes now clear that the first modes developing are those aperiodic instabilities at oblique propagation angles. For case 1 (top panel), the power spectrum of the a-PFI can reach values markedly higher than those of the p-PFI branch. But, again, this difference is less evident if electrons initially exhibit an anisotropy, e.g., in cases 2 and 3. This is in agreement with predictions from linear theory (see, for instance, the difference of maximum growth rates shown in Figure~\ref{gmax-r} and Figure~\ref{gmax} from Appendix~\ref{ApA}). 
In Figure~\ref{spectra_energy2} we can also observe that curves corresponding to the wave energy of the p-PFI (quasi-parallel) mode are not smooth, but exhibit an oscillatory, ripple-like evolution in time, more regular in Case~1 (when the initial electrons are isotropic). Similar profiles have been obtained in 1D simulations of p-EFI \citep{Messmer2002} and p-PFI \citep{Gary-etal-1998, Micera-etal-2020}. 
For the p-EFI, which are left-handed (LH) polarized, the oscillations of the magnetic wave energy seem to correspond to an oscillatory resonant heating of protons \citep{Messmer2002}, and suggests a similar explanation in our case.

\section{Discussion and conclusions}\label{conclusions}

It is thought that the Firehose instabilities (FIs) are important in constraining the anisotropy of electrons and ions in collision-poor plasmas from space, such as the solar wind. In particular, they are expected to play a role in the presence of anisotropy values $A=T_\perp/T_\parallel<1$, where perpendicular and parallel are intended with respect to the background magnetic field. These instabilities act to restore the isotropy.
Electrons (subscript $e$) and protons (subscript $p$) are dominant species in the solar wind. However, only relatively recent studies have proposed a realistic approach of the plasma states highly susceptible to these instabilities as conditioned by the interplay of both species when $A_e < 1$ and $A_p < 1$.
In the present paper, we have provided new insights from linear theory and the corresponding analysis from 2D semi-implicit PIC simulations, which enabled us to identify the fastest growing FIs near the instability thresholds and their long-term consequences on the relaxation of anisotropic distributions.

The unstable linear spectrum remains dominated by the aperiodic branch of EFI, i.e., a-EFI, predicted at oblique angles and with the highest growth rates, not much affected by the anisotropy of protons.
Instead, anisotropic electrons significantly influence the spectrum of PFIs at larger time scales. By comparing the maximum growth rate difference between aperiodic (a-PFI) and periodic (p-PFI) modes we have explored how the growth rates of these instabilities are modified. 
In contrast to previous studies, here we considered different, complementary regimes of PFIs that ensure favorable conditions for a first ignition of the a-PFI, propagating obliquely to the uniform magnetic field and developing faster than the p-PFI branch.  
We have compared the ideal case of isotropic electrons with that when electrons are anisotropic, with $A_e < 1$. 

Our PIC simulations confirm, in general, these theoretical predictions. 
While the a-EFI remains unaffected by the anisotropic protons, and saturates rapidly at low-level fluctuations, the subsequent PFI branches are markedly influenced by the anisotropic electrons. The a-PFI is excited before the p-PFI, and both branches are ignited much earlier. 
However, the competition between these two branches increases, since only p-PFI achieves a significantly higher fluctuating power. 
In return, the resulting enhanced
fluctuations of PFIs determine a deeper relaxation of the proton anisotropy.

Comparing the present results with the previous ones, we can conclude the following. 
Linear theory clearly shows that the regimes of electron-generated firehose instabilities (EFIs) do not undergo major changes, while PIC simulations confirm a general dominance of the a-EFI. 
Instead, the regimes of proton-triggered firehose instabilities (PFIs) are markedly changed, predicting greater effectiveness and competition of both branches, a-PFI and p-PFI, toward lower values of $\beta_p \leqslant 1$. 
One can also notice an extended dominance of the a-PFI at large deviations from isotropy $A_p \leqslant 0.1$ and high $\beta_p > 2$. 
The resulting fluctuations of PFIs are, in general, more robust than EFIs. But in the presence of anisotropic electrons with $A_e <1$, PFI fluctuations develop much earlier and saturate at higher levels of magnetic energy than obtained for isotropic electrons. 
Our results manage to clarify the specific conditions of the a-PFI mode, completing the previous analysis that left the impression of a universal dominance of the p-PFI mode.

\begin{acknowledgments}
R.A.L. acknowledges the support of ANID Chile through FONDECyT grant No.~11201048. 
The authors acknowledge support from the Katholieke Universiteit Leuven and Ruhr-University Bochum. These results were also obtained in the framework of the projects C14/19/089 (C1 project Internal Funds KU Leuven), G.0D07.19N (FWO-Vlaanderen), SIDC Data Exploitation (ESA Prodex-12), and Belspo project B2/191/P1/SWiM. A.M. and A.N.Z. thank the Belgian Federal Science Policy Office (BELSPO) for the provision of financial support in the framework of the PRODEX Programme of the European Space Agency (ESA) under contract Nos. 4000134474 and 4000136424. Powered@NLHPC: This research was partially supported by the supercomputing infrastructure of the NLHPC (ECM-02). Part of these simulations were performed on the supercomputers SuperMUC (LRZ) and Marconi (CINECA) under PRACE and HPC-Europa3 allocations.
\end{acknowledgments}

\appendix

\section{Growth rates for a reduced mass ratio} \label{ApA}

\begin{figure}[] 
\includegraphics[width=0.48\textwidth]{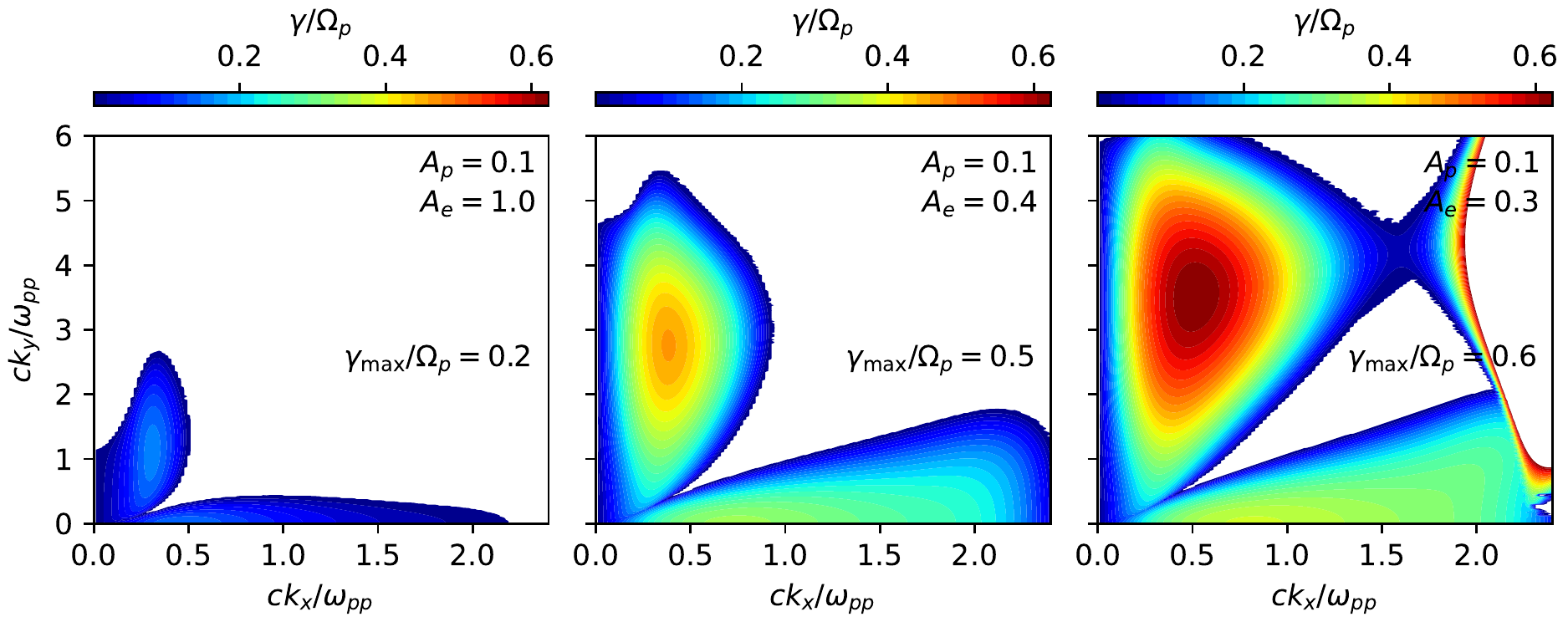}
\includegraphics[width=0.48\textwidth]{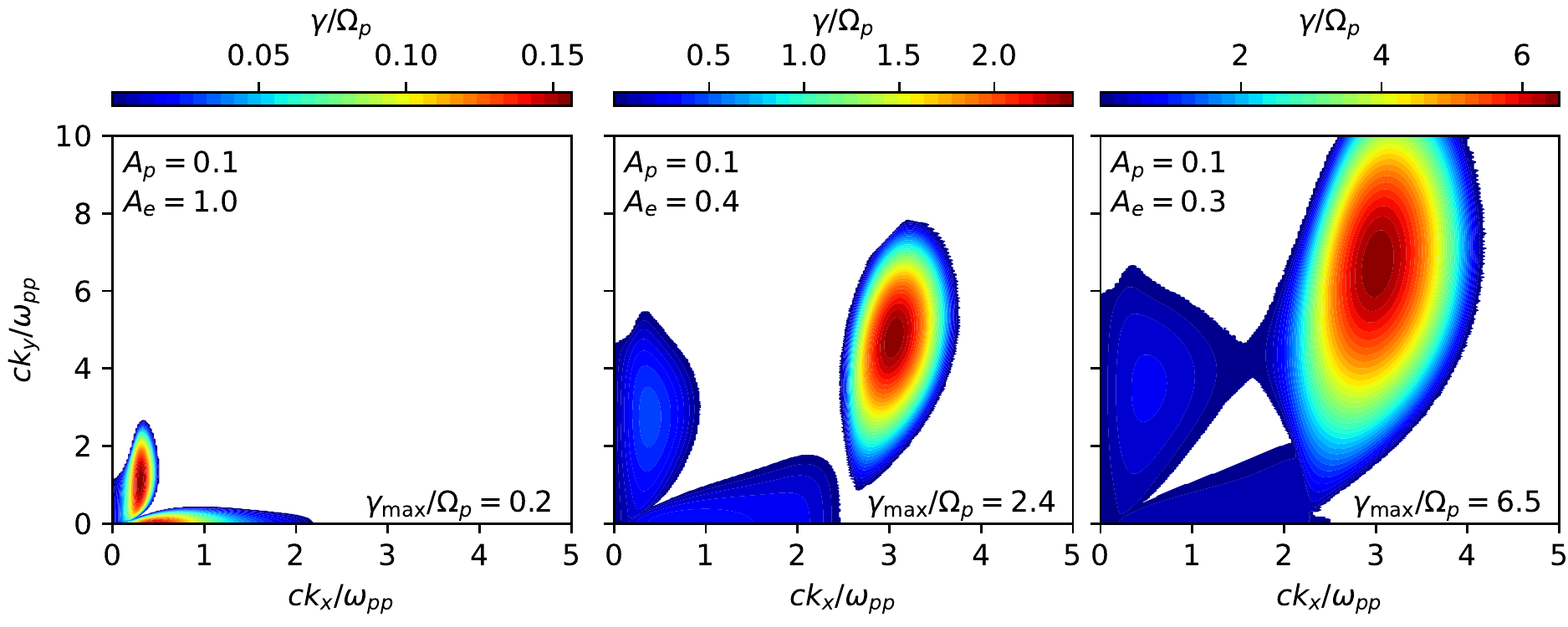}
\caption{Similar to Figure~\ref{pfhi-e-an} but for reduced mass ratio, $m_p/m_e=100$. Linear growth rate in the $k_x$--$k_y$ plane, for cases 1, 2 and 3 in Table~\ref{t1}. Left panels focus on PFI at smaller wave-numbers, while right panels show the complete spectra extended to larger wave-numbers.\label{disp_mpme}}
\end{figure}

Growth rates plotted in Figure \ref{disp_mpme} are obtained for the same parametric cases as in Figure~\ref{pfhi-e-an} (i.e., the same cases in Table \ref{t1}), except for a reduced mass ratio $m_p/m_e = 100$, as used in the simulations. 
In this case, the growth rate is computed as a function of $k_x$ and $k_y$ in order to compare with the spectra obtained from the simulation, as in Figures~\ref{spectra-EFI} and \ref{spectra-PFI}. 
On the other hand, from a comparison with Figure~\ref{pfhi-e-an} we can observe that only the maximum growth rates of the a-EFI branch are affected (which remain the highest, roughly, one order of magnitude higher than those of PFIs). In contrast, the maximum growth rate of PFIs, namely, that of a-PFI, practically does not change. Thus, a lower mass ratio reduces the differences between the maximum growth rates of PFIs and EFIs, but also between their specific ranges of unstable wavenumbers.

\begin{figure*}[t!] 
\centering
\includegraphics[width=0.3\textwidth]{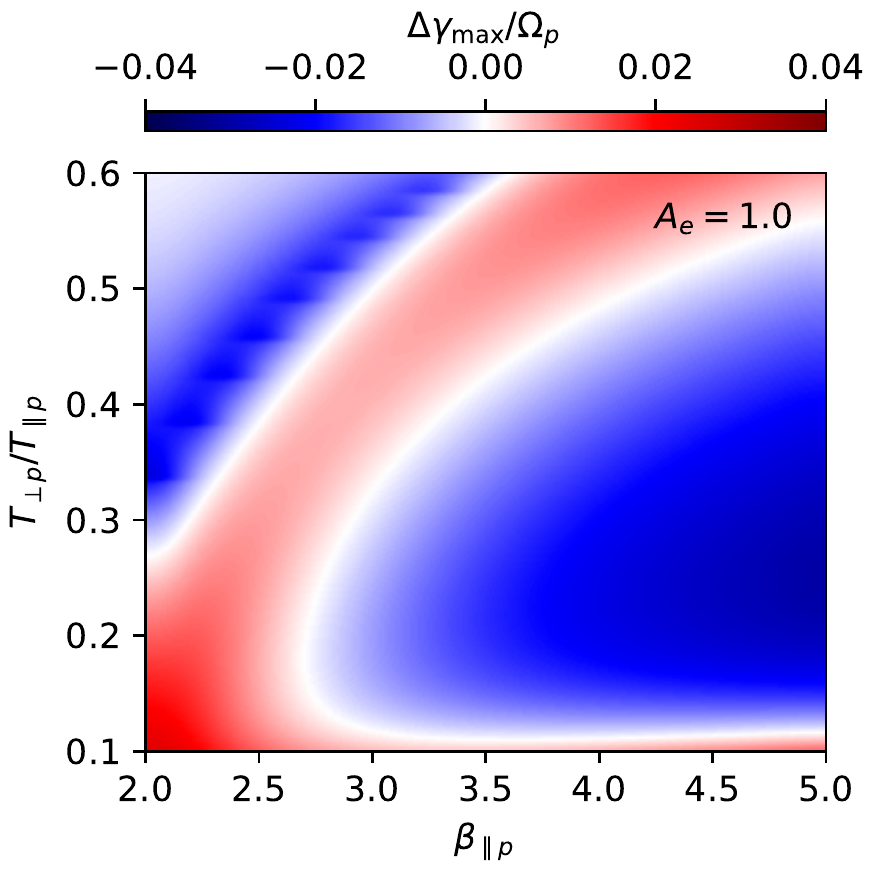}
\includegraphics[width=0.3\textwidth]{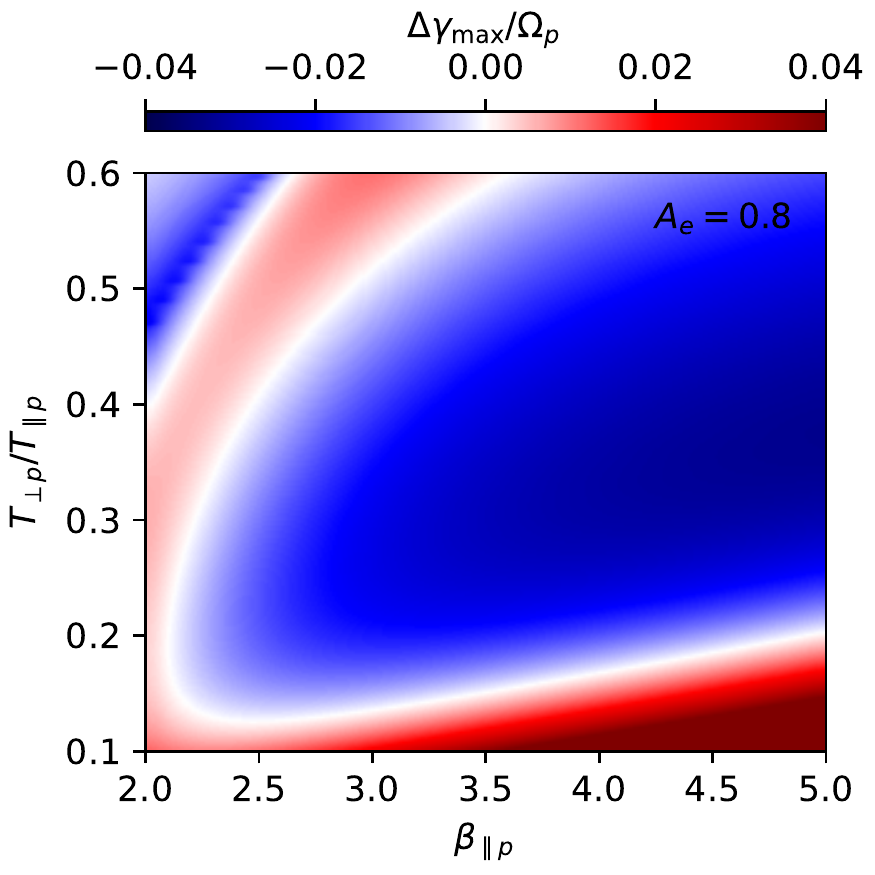}
\includegraphics[width=0.3\textwidth]{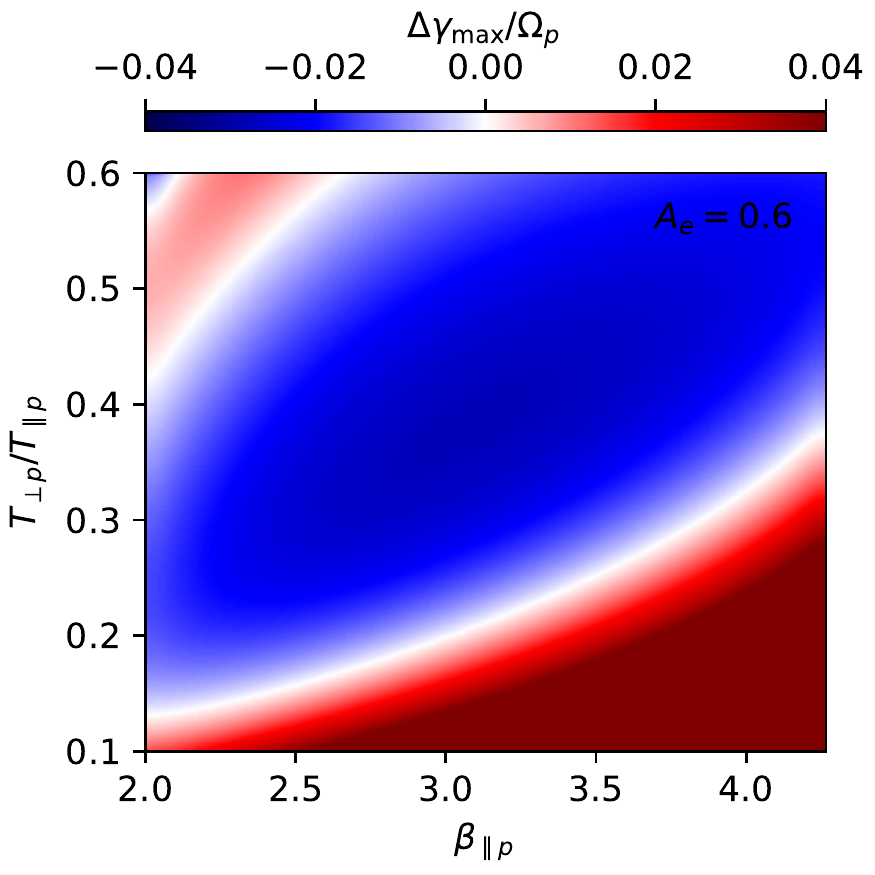}
\caption{As in Figure~\ref{gmax-r} but for reduced mass ratio, $m_p/m_e=100$. The difference between the maximum growth rates of a-PFI and p-PFI, $\Delta \gamma_\text{max} = \gamma_\text{max} (\text{a-PFI})-\gamma_\text{max}(\text{p-PFI})$, for $A_e=1$ (left), $A_e=0.8$ (center) and $A_e=0.6$ (right). Reddish shades indicate a dominance of a-PFI, while blue shades are regimes favorable to p-PFI.} \label{gmax}
\end{figure*}

Figure~\ref{gmax} displays the difference between the maximum growth rates of a-PFI and p-PFI, derived for an extended parametric domain including $2 \leqslant \beta_{p,\parallel} \leqslant 5.0$ and $0.2 \leqslant A_p \leqslant 0.6$. In this case the electron mass ratio is reduced to $m_p/m_e = 100$, and compared are the case of isotropic electrons $A_e = 1$ (left) with two situations with anisotropic electrons with $A_e = 0.8$ (middle) and 0.6 (right). Comparing with the results in Figure~\ref{gmax-r}, where these differences are derived for a realistic mass ration $m_p/m_e = 1836$, we can observe that the specific parametric regions where each of the a-PFI and p-PFI dominates, respectively, the red and blue regions, do not change significantly. Moreover, conditions chosen for cases 2 and 3 remain specific to a regime dominated by the a-PFI in the linear phase.

\listofchanges
\end{document}